%% file: main.tex
\documentclass[9pt,conference]{IEEEtran}
\IEEEoverridecommandlockouts
\usepackage{cite}
\usepackage{amsmath,amssymb,amsfonts}
\usepackage{circledsteps}
\usepackage{graphicx}
\usepackage{textcomp}
\usepackage{xcolor}
\usepackage{}
\usepackage[ruled,noline,noend]{algorithm2e}
\usepackage{fancyhdr}
\usepackage[normalem]{ulem}
\usepackage{flushend}
\usepackage{subfigure}
\usepackage{makecell}
\usepackage{booktabs}  
\usepackage{stfloats}
\usepackage{color}
\usepackage{graphicx}
\usepackage{amsmath}
\usepackage{setspace}
\usepackage{lipsum}
\usepackage{multirow}
\usepackage{multicol}
\usepackage{makecell}
\usepackage{soul}

\graphicspath{{./figures/}}

\newcommand{\ignore}[1]{}

\def\BibTeX{{\rm B\kern-.05em{\sc i\kern-.025em b}\kern-.08em
    T\kern-.1667em\lower.7ex\hbox{E}\kern-.125emX}}
\begin{document}

\title{ONE-SA: \underline{E}nabling \underline{N}onlinear \underline{O}perations in \underline{S}ystolic \underline{A}rrays For Efficient and Flexible Neural Network Inference}

\author{\IEEEauthorblockN{Ruiqi Sun$^{1}$, Yinchen Ni$^{1}$, Xin He$^{2}$, Jie Zhao$^{3}$, An Zou$^{1}$ \thanks{An Zou is the corresponding author. This work is supported by NSFC 62202287.}}
$^{1}$Shanghai Jiao Tong University, $^{2}$University of Michigan, $^{3}$Microsoft Corporation
}

\maketitle

\begin{abstract}
Deep neural networks (DNNs) have achieved significant success in wide fields, such as computer vision and natural language processing. However, their computation and memory-intensive nature limits their use in many mobile and embedded contexts. Application-specific integrated circuit (ASIC) hardware accelerators employ matrix multiplication units (such as the systolic arrays) and dedicated nonlinear function units to speed up DNN computations. A close examination of these ASIC accelerators reveals that the designs are often specialized and lack versatility across different networks, especially when the networks have different types of computation. In this paper, we introduce a novel systolic array architecture, which is capable of executing nonlinear functions. By encompassing both inherent linear and newly enabled nonlinear functions within the systolic arrays, the proposed architecture facilitates versatile network inferences, substantially enhancing computational power and energy efficiency. Experimental results show that employing this systolic array enables seamless execution of entire DNNs, incurring only a negligible loss in the network inference accuracy. Furthermore, assessment and evaluation with FPGAs reveal that integrating nonlinear computation capacity into a systolic array does not introduce extra notable (less than 1.5\%) block memory memories (BRAMs), look-up-tables (LUTs), or digital signal processors (DSPs) but a mere 13.3\% - 24.1\% more flip flops (FFs). In comparison to existing methodologies, executing the networks with the proposed systolic array, which enables the flexibility of different network models, yields up to 25.73$\times$, 5.21$\times$, and 1.54$\times$ computational efficiency when compared to general-purpose CPUs,  GPUs, and SoCs respectively, while achieving comparable (83.4\% - 135.8\%) performance with the conventional accelerators which are designed for specific neural network models.
\end{abstract}


\vspace{-5mm}

\input{introduction}
\input{background_and_relatedworks}
\input{Design}
\input{Optimization}

\input{Evaluation}
\input{Conclusion}
\begin{spacing}{0.9}
\bibliographystyle{unsrt}
\bibliography{bibliography}
\end{spacing}

\end{document}

%% file: introduction.tex
\section{Introduction}
\label{sec:introduction}
\begin{figure}
\setlength{\abovecaptionskip}{-0.02cm}
\begin{minipage}[t]{0.49\linewidth} 
\centering
\subfigure[CNN-based ResNet]
{\includegraphics[trim={0cm 0cm 1cm 0cm},clip,width=0.99\linewidth]{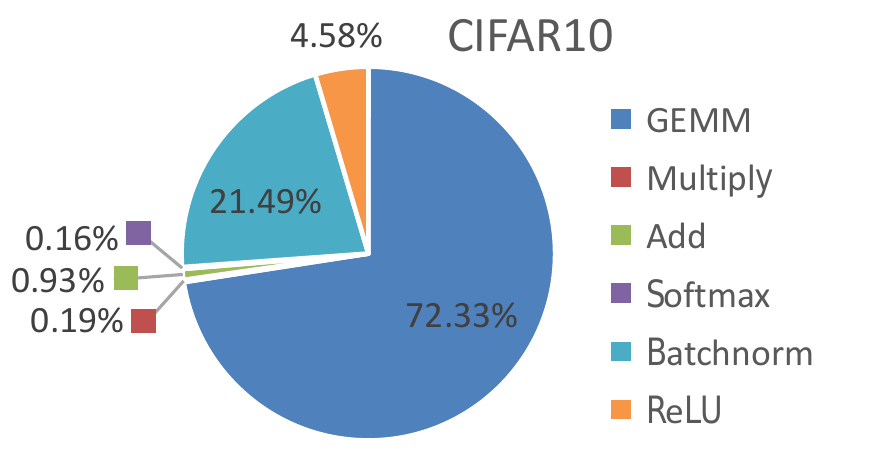}} 
\end{minipage}%
\begin{minipage}[t]{0.416\linewidth}
\centering
\subfigure[Transformer-based BERT]
{\includegraphics[trim={0cm 0cm 1cm 0cm},clip,width=0.99\linewidth]{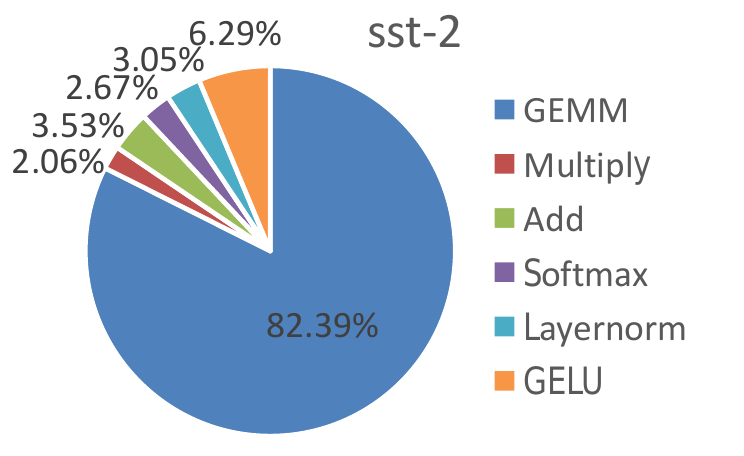}}
\end{minipage}%
\caption{The computations in classic neural network models.}
\label{fig:computation}
\vspace{-7mm}
\end{figure}

Deep neural networks (DNNs) have achieved remarkable success across diverse domains, including computer vision (CV) and natural language processing (NLP). However, their widespread deployment faces significant challenges due to the substantial computational demands and high power consumption associated with DNNs. These limitations restrict their use on resource-constrained platforms. The computational complexity of DNN models encompasses both linear calculations and essential nonlinear operations. In a general context, linear computations can be succinctly expressed as general matrix multiplications. However, nonlinear operations, which have been proven to be a must in neural networks \cite{eckle2019comparison}, exhibit versatility and often align with specific characteristics of individual network models, as presented in Fig. \ref{fig:computation}.

Application-specific integrated Circuits (ASICs) offer a promising solution for empowering DNN inferences on mobile and edge devices. These ASIC accelerators, which integrate a range of functional units, including matrix multiplication units for linear computations and dedicated units (GELU, Softmax, etc.) for nonlinear functions, streamline the intricate computations inherent to neural networks. For instance, an edge-based BERT accelerator~\cite{tambe2021edgebert} incorporates specialized processing units equipped with Floating-Point (FP) vectors and accumulate capabilities, as well as dedicated function units for each nonlinear operation such as Layer normalization and Softmax calculations. However, upon closer examination of these ASIC-based accelerator designs, several common challenges emerge that warrant further attention. The inherent specificity of ASIC-based accelerators, although beneficial for targeted applications, often lack the flexibility to efficiently handle a broader range of DNN models. The distinct data flow patterns from various buffers to diverse computing units can lead to substantial performance stalls~\cite{khan2021npe}. Moreover, unless meticulously designed with perfect data flow and pipelining, one computing unit may remain idle while another processes the workload.

The systolic array is a parallel computing architecture often utilized within ASIC accelerators. It is featured by a regular and structured design, optimizing the efficiency of repetitive linear mathematical computations, such as general matrix multiply and convolution \cite{choi2023enabling}.
This paper introduces an architectural design for systolic arrays called ONE-SA (meaning One-size-fits-all Systolic Array), targeting to support a wide range of nonlinear computations. By combining this capability with the inherent efficiency of linear computations, the proposed architecture enables the seamless execution of versatile computations required by DNNs through a single computing unit. This architecture exhibits the flexibility of accommodating various DNN models efficiently. Another advantage of ONE-SA lies in its ability to maintain continuous computation, eliminating idle periods often associated with imperfect pipeline designs. The implementation and evaluation with FPGA demonstrate that ONE-SA achieves computational and power efficiency levels that were only attainable by the application-specific accelerators, while it retains the flexibility to support a wide range of DNN models.
The contributions of this paper can be summarized as follows: \vspace{-4mm}

\begin{itemize}
\item A systolic array architecture is introduced to facilitate nonlinear operations, which are initially approximated through capped piecewise linearization and then computed by Matrix Hadamard Products (MHPs) with intermediately fetched parameters.

\item The micro-architectures of function units, such as the processing element and L3 buffer, are meticulously designed considering the tradeoffs between computation performance and hardware resource cost when performing computations.

\item A series of algorithm experiments and FPGA implementations demonstrated that ONE-SA achieves efficiency levels exclusive to state-of-the-art ASIC-based accelerators. Furthermore, the efficiency and flexibility of ONE-SA is improved in executing a wide range of DNN models. 

\end{itemize}

%% file: background_and_relatedworks.tex
\section{Background and Related Work}
\label{sec:background_and_relatedworks}

\begin{figure}[!t]
    \setlength{\abovecaptionskip}{-0.04cm}
    \centering
    \includegraphics[width = 0.28\textwidth]{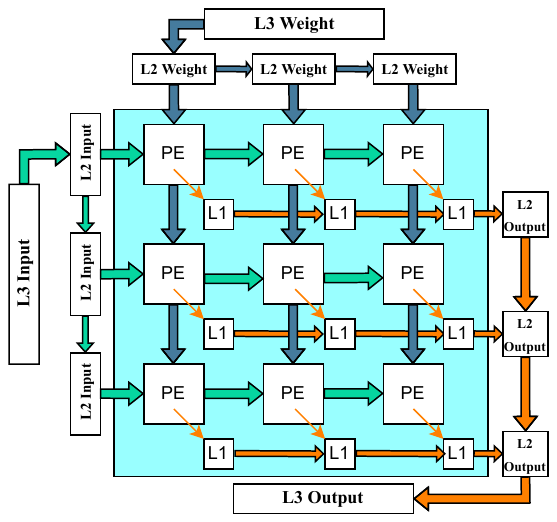}
    \caption{A classic systolic array architecture with 3x3 PEs.}
    \label{fig:SA_3_3}
    \vspace{-6mm}
\end{figure}
\subsection{Systolic Arrays for Linear Compuataions}
Systolic arrays, also called Matrix Processing Units (MXU) or General Matrix Multiplication Accelerators,  constitute a specialized hardware architecture tailored for parallel linear computation, especially in tasks involving matrix multiplications, a foundational operation in deep neural networks \cite{wei2017automated}. As depicted in Fig. \ref{fig:SA_3_3}, the systolic array features a grid of processing elements (PEs) and memory hierarchy (three levels of buffers) to execute computations in parallel, allowing for intensive parallel calculations \cite{wang2021autosa}. Each PE can process multiply-accumulate operations and send/receive data to/from its adjacent PEs or buffers. Systolic arrays are adept at efficiently managing input data and weight sharing and can be scaled to align with the computation demands, often mitigating memory (i.e., buffers) bandwidth requirements. Benefiting the performance and efficiency in linear computation, systolic arrays play a central role in running DNN tasks \cite{wei2017automated}. While systolic arrays excel at supporting linear operations, which can be expressed as general matrix multiplication (like im2col-based convolution), nonlinear operations have been demonstrated to be essential components in the neural network and cannot be effectively computed by systolic arrays. Consequently, in neural network accelerators, specialized function units like activation units and normalization/pooling units are integrated alongside systolic arrays to handle these nonlinear operations \cite{7930521}. The types and quantities of nonlinear functions required vary for each network model. As a result, the accelerator equipped with a systolic array and application-specific nonlinear function units can achieve computation and power efficiency but must be tailored to specific network models.
\vspace{-5mm}
\subsection{Simplification and Approximation for Nonlinear Operations}
\vspace{-1mm}
Non-linear operations such as GELU, Layer normalization, and Softmax are essential yet costly building blocks for neural networks like convolutional neural networks (CNNs) and transformers. Besides, the straightforward dealing with the nonlinear operations with the specific nonlinear function units, several prior works simplified the nonlinear operations with all integer computations \cite{kim2021bert} or matrix computation \cite{ming2022ma}, but such simplification mainly focuses on a specific algorithm and tested on the general-purpose processors such as CPUs and GPUs. Recently, approximations such as the piecewise linear (PWL)-based approximation \cite{dong2020plac, khan2021npe, lyu2021ml} and the neural network-based approximation \cite{yu2022nn} are also proposed to accelerate the nonlinear operations in neural networks. To simplify and automate the approximation progress for neural networks, an automated approximation framework \cite{lu2023auto}, which leverages a neural network to automatically approximate non-linear operations, is developed by researchers. However, most approximation approaches still require additional and separate computing and storage units alongside the systolic arrays to finish the approximation computing. This paper proposes a systolic array architecture that enables the nonlinear function in the systolic array. Together with the inherent support of linear computation, the proposed systolic array can achieve efficient and flexible execution with versatile neural network models.

%% file: Design.tex
\section{Architecture Design}
\label{sec:architecture}
In this section, we propose an architecture design for approximating nonlinear functions using capped piecewise linearization (CPWL) and the approximation is then calculated through Intermediate Parameter Fetching (IPF) and Matrix Hadamard Product (MHP).

\subsection{From Nonlinear Operations to IPF and MHPs}
\begin{figure}
\setlength{\abovecaptionskip}{-0.04cm}
    \centering
    \includegraphics[width = 0.48\textwidth]{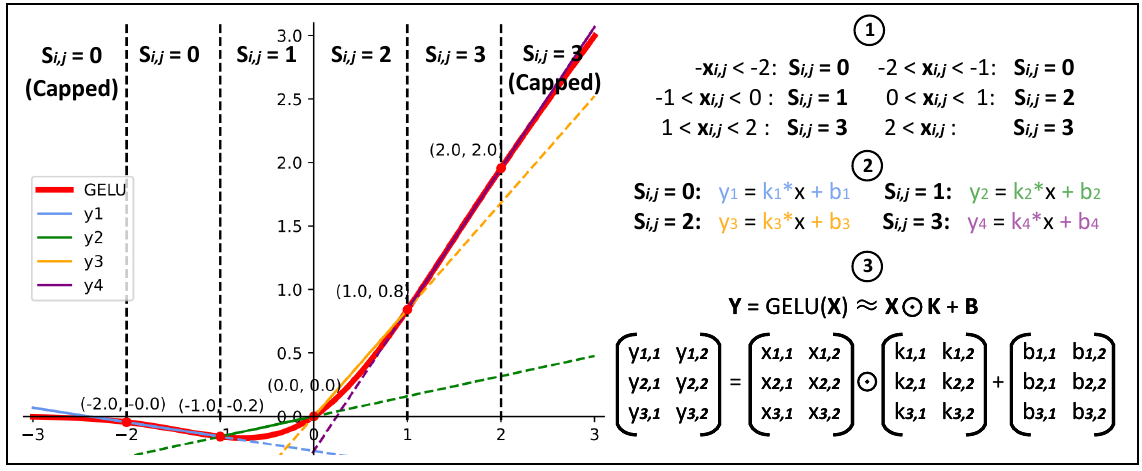}
    \caption{Capped piecewise linearization for nonlinear operations.}
    \label{fig:overview}
    \vspace{-5mm}
\end{figure}

In ONE-SA, continuous nonlinear operations are approximated using CPWL. The CPWL approximation offers two primary advantages over alternative methods, such as Taylor expansion or Chebyshev approximation. Firstly, CPWL entails only linear computation, readily accommodated by the computation circuits within the processing elements (PEs), whereas alternative approximations necessitate additional computational circuitry. Secondly, computations in CPWL are simple. A group of CPWL can be efficiently parallelized, capitalizing on the proven efficiency of systolic arrays for parallel processing.

Here, we use the nonlinear function GELU as an illustrative example (in Fig. \ref{fig:overview}), but the same process can also be used to handle other nonlinear operations, such as Softmax and Layer normalization. 
In the capped piecewise linear approximation, the original nonlinear function $y=\text{GELU}(x)$ is cut into many small segments according to the input value $x$. In each segment, the original nonlinear function $y=\text{GELU}(x)$ is approximated with $y=kx+b$. The parameters $k$ and $b$ are the slope and bias of the line connecting the start and end points of the segment noted as $s$. Since the nonlinear function is known, the segment $s$ and parameters $k$ and $b$ for each segment can be pre-calculated given approximation granularity (segment length).
Then, a three-step approach is presented to realize the calculations for a group of parallel nonlinear operations: $Y=\text{GELU}(X) \approx X \odot K+B$. The $X, Y \in \mathbb R^{M\times N}$ are the group of parallel input $x_{i,j}$ and output $y_{i,j}$ expressed with a matrix format. The pre-calculated parameters $k$ and $b$ of each segment are pre-stored and indexed by the segment numbers.
The capped piecewise linear calculation follows the following steps: \textbf{\scriptsize \Circled{1}} Calculate the segment matrix $S$ for the input matrix $X$. Each element in $S$ (i.e., $s_{i,j}$), represents which segment its corresponding input value $x_{i,j}$ falls into. If the value of $s_{i,j}$ exceeds the valid range of segments, the value of $S_{i,j}$ will be capped, and the values of the boundary segment will be given to $s_{i,j}$.
\textbf{\scriptsize \Circled{2}} The intermediate parameters $k$ and $b$ are aggregated and sent to the systolic array in the forms of slope matrix $K$ and intercept matrix $B$; 
\textbf{\scriptsize \Circled{3}} Finally, the element-wise calculations $Y = X \odot K+B$ are performed to get the output matrix $Y$. 
To support the three-step approach in the systolic array, we categorize it into two architecture level events: Intermediate Parameter Fetching (IPF) for the steps \textbf{\scriptsize \Circled{1}} and \textbf{\scriptsize \Circled{2}} and Matrix Hadamard Product (MHP) for the step \textbf{\scriptsize \Circled{3}}.

\begin{figure}
\setlength{\abovecaptionskip}{-0.1cm}
    \centering
    \includegraphics[width = 0.42\textwidth]{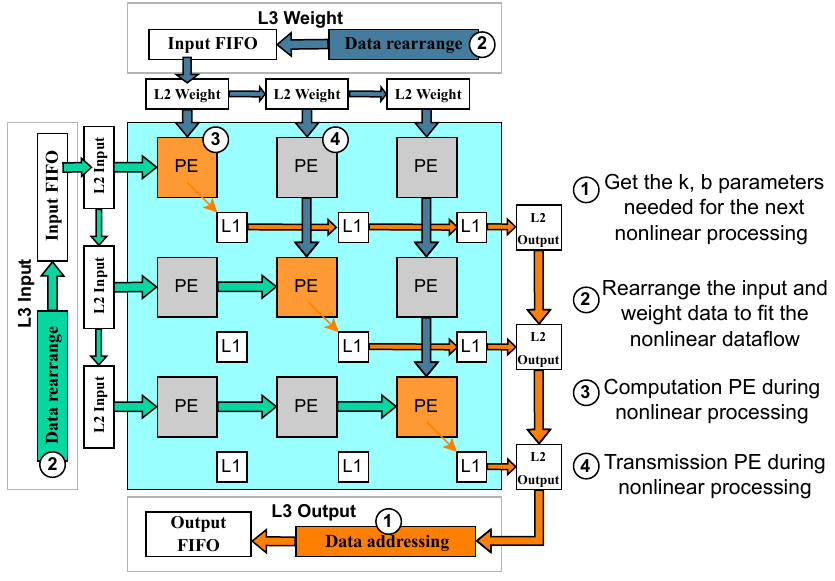}
    \caption{Proposed systolic array architecture with nonlinear operations.}
    \label{fig:SA_3_3_nonlinear}
    \vspace{-5mm}
\end{figure}

\subsection{Architecture Support for IPF and MHPs}
The new systolic array architecture should meet the following design criteria: First, functionally supporting Intermediate Parameter Fetching (IPF) and Matrix Hadamard Product (MHP). Second, facilitating a high degree of computational performance in newly enabled nonlinear operations. Third, maintaining a compact design and no harm to inherent classic linear operation (i.e., general matrix multiply).
As systolic arrays find common use in general-purpose matrix multiply, they can naturally provide the computation units used for MHP and further capped linear approximation. 
However, there lacks an efficient path to fetch the intermediate parameters ($k$ and $b$), according to the inputs, to the PEs. Meanwhile, direct execution of these  MHPs exhibits a low resource utilization rate as the conventional data reuse will be discarded in the MHP.
The architecture of the proposed systolic array is illustrated in Fig. \ref{fig:SA_3_3_nonlinear}. The fundamental concept behind this design comprises four essential microarchitecture improvements. \textbf{\scriptsize \Circled{1}} \textbf{Data Addressing} and \textbf{\scriptsize \Circled{2}} \textbf{Data Rearrange and Fetch} develop an efficient data path for retrieving and distributing intermediate parameters for the following MHP. \textbf{\scriptsize \Circled{3}} \textbf{Computation PE} and \textbf{\scriptsize \Circled{4}} \textbf{Transmission PE} build a novel data path within each PE, facilitating efficient data flow for MHPs, where the values in input and weight matrix are used only once.
\vspace{-2mm}

%% file: Optimization.tex
\section{Microarchitecture Design and Optimization}
\vspace{-1mm}
\label{sec:Optimization}
This section will dive into the microarchitecture design and optimization of each functional module that supports the implementation of nonlinear functions through Intermediate Parameter Fetching (IPF) and Matrix Hadamard Product (MHP). It is crucial to note that these microarchitectural design and optimization considerations should not compromise the performance of the general matrix multiply, which is a fundamental prerequisite.

\vspace{-1mm}
\subsection{L3 Buffer for Intermediate Parameter Fetching}
As introduced in Section \ref{sec:architecture}, the capped piecewise linear approximation necessitates the use of intermediate parameters $k$ and $b$ with a matrix format $K, B \in \mathbb R^{M\times N}$.
Before intermediate parameters can be used in the Matrix Hadamard Product, they should be addressed according to the value of each element $x_i$ in the input matrix $X$. To support the addressing and delivery of parameters $k$ and $b$ with a matrix format $K, B \in \mathbb R^{M\times N}$, we design additional modules within the L3 buffer. 

\begin{figure}[b]
    \setlength{\abovecaptionskip}{-0.1cm}
    \centering
    \includegraphics[width = 0.28\textwidth]{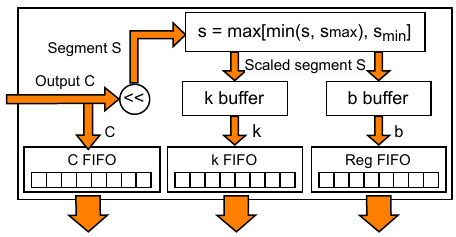}
    \caption{The microarchitecture of L3 buffers data addressing.}
    \label{fig:L3_buffer}
    \vspace{-5mm}
\end{figure}

\begin{figure}
\setlength{\abovecaptionskip}{-0.1cm}
    \centering
    \includegraphics[width = 0.35\textwidth]{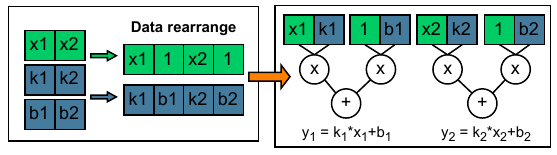}
    \caption{The microarchitecture of data rearrangement.}
    \label{fig:data_rearrange}
    \vspace{-5mm}
\end{figure}

\subsubsection{Data Addressing}
The data addressing module is depicted in Fig. \ref{fig:L3_buffer} and comprises several key components, including a data shift module, a scale module, two buffers, and three FIFOs. This module enables the seamless extraction and storage of $k$ and $b$ values simultaneously. The output $S$ (the previous $C$ used in the general matrix multiply) is loaded with input matrix $X$, and further, the element $x_{i,j}$ in the input matrix $X$ is fed into the data shift module alongside the main output. The data shift module's role is to calculate the segment number $s_{i,j}$, leveraging the efficiency of data shifting when segment lengths are set to powers of two. Subsequently, the calculated segment number $s_{i,j}$ is forwarded to the scale module, which intervenes in case the segment number exceeds the predefined range for preload parameters. The scaled segment number then serves as an address to retrieve the preload $k$ and $b$ parameters, which are loaded to the L3 buffer simultaneously with the input $X$ before. Finally, the $k$ and $b$ parameters are extracted and written to DRAM. This design ensures parameters $k$ and $b$ behave like the conventional output $C$ in general matrix multiply in the DRAM, ready for the following Matrix Hadamard Product.

\subsubsection{Data Rearrange and Fetch}
The Matrix Hadamard Product involves three metrics: $X$, $K$, and $B$. However, the conventional systolic array is equipped with only two input channels. Adding an additional channel would result in increased hardware resource costs and decreased hardware utilization rates. Therefore, it is imperative to merge matrices $K$ and $B$ into a single matrix prior to input.
In our design, after the intermediate data addressing, the intermediate parameters $k$ and $b$ are retrieved from DRAM and sent to the memory relocation module. Here, we meticulously reorganize the data, consolidating each element of $k$ with its corresponding $b$ to form a new data stream, as shown in Fig. \ref{fig:data_rearrange}. Simultaneously, we employ a similar approach to process the matrix $X$, where each element of $X$ is paired with the number 1 to create a new data stream. This preparation ensures that the PE can accurately execute piecewise linear computations. This efficient configuration guarantees the accurate execution of the capped piecewise linear calculations while maintaining low hardware resource costs.\vspace{-2mm}
\begin{figure}[h]
\setlength{\abovecaptionskip}{-0.1cm}
\begin{minipage}{0.33\linewidth} 
\vspace{-3mm}
\centering
\subfigure[Overall architecture]
{\includegraphics[trim={0cm 0cm 0cm 0cm},clip,width=0.91\linewidth]{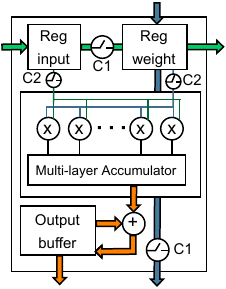}} 
\end{minipage}%
\begin{minipage}{0.33\linewidth}
\vspace{-3mm}
\centering
\subfigure[Computation PE]
{\includegraphics[trim={0cm 0cm 0cm 0cm},clip,width=0.91\linewidth]{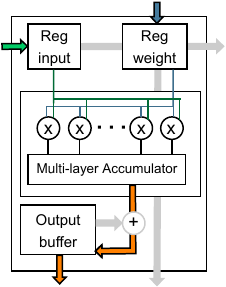}}
\end{minipage}%
\begin{minipage}{0.33\linewidth}
\vspace{-3mm}
\centering
\subfigure[Transmission PE]
{\includegraphics[trim={0cm 0cm 0cm 0cm},clip,width=0.91\linewidth]{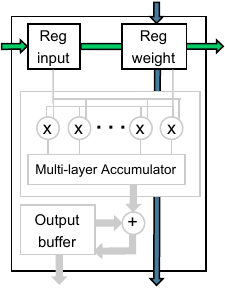}}
\end{minipage}%
\caption{The microarchitecture PEs in the proposed ONE-SA.}
\label{fig:microarchitecture_pe}
\vspace{-5mm}
\end{figure}
\subsection{Processing Element for Matrix Hadamard Product}
As discussed in the previous section, it is evident that each element must execute a multiply-accumulate (MAC) operation with the segment length or specific parameters $k$ and $b$ during the piecewise linear approximation process. This process is characterized by a low data reuse rate, primarily because each element possesses unique parameters that cannot be shared with other elements. In a conventional systolic array architecture, the PEs typically receive input and weight data from preceding PEs, perform a MAC operation, and subsequently transmit the modified weight and input data to the next PE both horizontally and vertically. However, this architectural design, which is optimized for high data reuse rates in the general matrix multiply, is ill-suited for handling the demands of the MHP due to its inherently low data reuse characteristics. Therefore, improving the microarchitecture of PE becomes imperative, focusing on enabling it to efficiently accommodate computations with low data reuse rates without hurting its capacity for high data reuse when processing general matrix multiply operations.

\subsubsection{Data Flow}
Our initial step involves the design of the dataflow for the MHP, as illustrated in Fig. \ref{fig:microarchitecture_pe}. As previously mentioned, it is essential to recognize that each element's input and weight data are unique and cannot be shared with other elements. Consequently, the input and weight data transmitted in both directions of the systolic array can only undergo computation once and cannot be further forwarded to the next PE after this computation. This leads us to the conclusion that only PEs positioned along the diagonal can actively engage in calculations during the MHP computation process, while the remaining PEs essentially serve as data registers. Our design categorizes PEs into two distinct types: computation PEs and transmission PEs. The data passed through the transmission PEs within each row and column and routed to the computation PEs located on the diagonal. Once these data have been computed, they are no longer required for subsequent processing.

\subsubsection{Design Components}
To implement the proposed dataflow, we build a new path in the PE, as depicted in Fig. \ref{fig:microarchitecture_pe} (a). This enhanced PE features four additional Control Logics (noted as C1 and C2) and related auxiliary circuitries. In contrast to the conventional PE module, these modifications provide the flexibility to select different functions.
During the execution of general matrix multiply operations, both Control Logic C1 and C2 are activated, enabling the PE to function as a conventional systolic array PE. Concurrently, the multi-layer accumulator accumulates the results from parallel multiplication units and subsequently writes the final output to the buffer.
A reconfiguration is employed in the context of MHP computations. PEs located along the diagonal (i.e., Computation PE) have Control Logic C1 deactivated and Control Logic C2 activated. This configuration ensures that the received input and weight data are computed locally without transmission to subsequent PEs. The multi-layer accumulator directly writes the first layer's result to the output buffer, as depicted in Fig. \ref{fig:microarchitecture_pe} (b).
For the remaining PEs (i.e., Transmission PE), Control Logic C1 remains active, while Control Logic C2 is deactivated. This configuration allows the received input and weight data to be transmitted to subsequent PEs without undergoing local computation, as illustrated in Fig. \ref{fig:microarchitecture_pe} (c). \vspace{-2mm}

\begin{table}\scriptsize
  \centering
  \caption{Resources consumption of the ONE-SA L3 and PE.\vspace{-2mm}}
    \begin{tabular}{c|c|cccc}
    \hline
    Module & Design & BRAM  & LUT   & FF    & DSP \\
    \hline
        \multirow{2}[2]{*}{L3} & SA    & 0     & 174   & 566   & 0 \\ 
          & ONE-SA & 2     & 1021  & 1209  & 0 \\
    \hline
    \multirow{2}[2]{*}{PE} & SA    & 1     & 824   & 1862  & 16 \\ 
          & ONE-SA & 1     & 826   & 2380  & 16 \\
    \hline
    \end{tabular}%
  \label{tab:PE_L3}%
  \vspace{-6mm}
\end{table}%

\begin{table}\scriptsize
  \centering
  \caption{Total hardware resources consumption comparison.\vspace{-2mm}}
    \begin{tabular}{c|c|cccc}
    \hline
    Dim   & Design & BRAM  & LUT   & FF    & DSP \\
    \hline
    \multirow{3}[4]{*}{4*4} & SA    & 470   & 67976 & 66924 & 256 \\
\cline{2-6}          & \multirow{2}[2]{*}{OneSA} & 472   & 68855 & 75855 & 256 \\
          &       & (100.4\%) & (101.3\%) & (113.3\%) & (100\%) \\
    \hline
    \multirow{3}[3]{*}{8*8} & SA    & 822   & 179247 & 179247 & 1024 \\
\cline{2-6}          & \multirow{2}[1]{*}{OneSA} & 824   & 180222 & 213042 & 1024 \\
          &       & (100.2\%) & (100.5\%) & (118.9\%) & (100\%) \\
    \hline
    \multirow{3}[3]{*}{16*16} & SA    & 1366  & 730225 & 552539 & 4096 \\
\cline{2-6}          & \multirow{2}[2]{*}{OneSA} & 1368  & 731584 & 685790 & 4096 \\
          &       & (100.1\%) & (100.2\%) & (124.1\%) & (100\%) \\
    \hline
    \end{tabular}%
  \label{tab:resource_cost}%
  \vspace{-4mm}
\end{table}%

\begin{table}[tbp]\scriptsize
\vspace{-1mm}
  \centering
  \caption{End-to-end inference accuracy of different DNN models with different tasks.\vspace{-2mm}}
  \begingroup
  \setlength{\tabcolsep}{5pt}
    \begin{tabular}{cl|cccccc}
    \hline
    DNN   & Dataset & Original  & 0.1   & 0.25  & 0.5   & 0.75  & 1 \\
    \hline
    \multirow{4}[2]{*}{CNN} & QMNIST & 100.0\% & 0.0\% & 0.0\% & 0.0\% & -0.4\% & -0.7\% \\
          & Fashion-M & 91.2\% & -0.2\% & -0.4\% & -1.5\% & -2.5\% & -3.9\% \\
          & CIFAR-10 & 96.2\% & -0.8\% & -2.2\% & -3.0\% & -4.1\% & -5.3\% \\
          & CIFAR-100 & 85.1\% & -1.3\% & -2.8\% & -3.7\% & -6.3\% & -8.9\% \\
    \hline
    \multirow{4}[2]{*}{BERT} & SST-2 & 92.3\% & -0.1\% & -0.2\% & -0.4\% & -1.1\% & -2.2\% \\
          & QNLI  & 90.7\% & -0.9\% & -1.2\% & -1.7\% & -2.3\% & -2.7\% \\
          & STS-B & 88.7\% & -2.1\% & -2.7\% & -4.6\% & -6.6\% & -7.6\% \\
          & CoLA  & 56.5\% & -0.4\% & -1.2\% & -3.0\% & -11.3\% & -17.9\% \\
    \hline
    \multirow{4}[2]{*}{GCN} & Reddit & 92.7\% & -0.2\% & -0.6\% & -0.9\% & -1.2\% & -1.5\% \\
          & CORA  & 84.3\% & -0.2\% & -0.2\% & -0.2\% & -0.3\% & -4.3\% \\
          & Pubmed & 74.5\% & -0.1\% & -0.3\% & -0.2\% & 0.0\% & -0.2\% \\
          & Citeseer & 64.6\% & -1.8\% & -1.9\% & -1.8\% & -1.9\% & -2.8\% \\
    \hline
    \end{tabular}%
  \endgroup
  \label{tab:accuracy}%
  \vspace{-6mm}
\end{table}%

\subsection{Optimization of Critical Path Delay with Resource Cost}
\vspace{-1mm}
Optimizing the critical path typically involves striking a balance between enhancing performance to meet timing constraints and effectively managing hardware resource utilization. From the synthesis, the buffers easily become the critical paths. During the design, we prefer to allocate more resources to reduce the delays in the buffers.
As the FPGA implementation described in Section \ref{sec:Evaluation}, Table \ref{tab:PE_L3} presents the resource utilization data for the proposed PE and L3 buffer. when implemented on the FPGA. 
In comparison to traditional systolic arrays, the proposed PE exhibits identical BRAM and DSP consumption, nearly equivalent LUT utilization, and a 27\% increase in FFs. This additional FF usage is primarily attributed to Control Logic components.
Since the L3 buffer plays a pivotal role in determining the critical path within the ONE-SA architecture, we are inclined to allocate extra resources to optimize this path for performance gains. In contrast to conventional systolic arrays, the proposed L3 buffer necessitates 4.87x more LUTs and 1.14x more FFs. Despite the relatively more allocation of LUTs and FFs to the L3 buffer, the absolute numbers remain comparable to or even less than those within a single PE. Given the systolic array's topology, where there are multiple PEs compared to a single L3 buffer, it is justifiable to allocate additional resources to the L3 buffer in exchange for critical path optimization. Table \ref{tab:resource_cost} summarizes the comprehensive resource utilization across various systolic array sizes. The innovative ONE-SA design introduces a modest increase in FFs composition, ranging from 13.3\% to 24.1\%, while maintaining almost the same consumption for BRAMs, LUTs, and DSPs. Furthermore, it's worth noting that FFs are relatively less costly toward memory or computational functions compared to BRAMs, LUTs, and DSPs. \vspace{-3mm}

%% file: Evaluation.tex
\section{Implementation and Evaluation}
\label{sec:Evaluation}
\vspace{-2mm}
\subsection{Implementation and Experimental Setup}
\vspace{-1mm}
We implement the proposed systolic array architecture design ONE-SA with the Xilinx Virtex 7 XC7VX485T FPGA. We converted C/C++ code to RTL implementation using Vivado HLS 2019.2. The power consumption is reported from Xilinx Power Estimator (XPE) \cite{XPDUG}. The design of auxiliary circuitry, like the logic control units and the bandwidth of memory and buses, follows the high-performance systolic array design in \cite{wang2021autosa}.
we adopt some of the most well-known pre-trained DNNs in each category as the candidate network to test the proposed ONE-SA, such as ResNet-50 \cite{he2016deep}, BERT-base \cite{devlin2018bert}, and GCN \cite{kipf2016semi}, to represent CNN, transformer, and GNN, respectively. To ensure a comprehensive evaluation, both the neural networks and the systolic arrays are quantized to INT16 precision. The network inference accuracy, the performance and cost of the proposed systolic array, and the comparison with the general-purpose and application-specific processors are presented.

\begin{figure}
\begin{center}

\setlength{\abovecaptionskip}{-0.1cm}
\subfigure[Linear Calculation]{
    \includegraphics[trim=0cm 0cm 0cm 0cm, clip,width=0.235\textwidth]{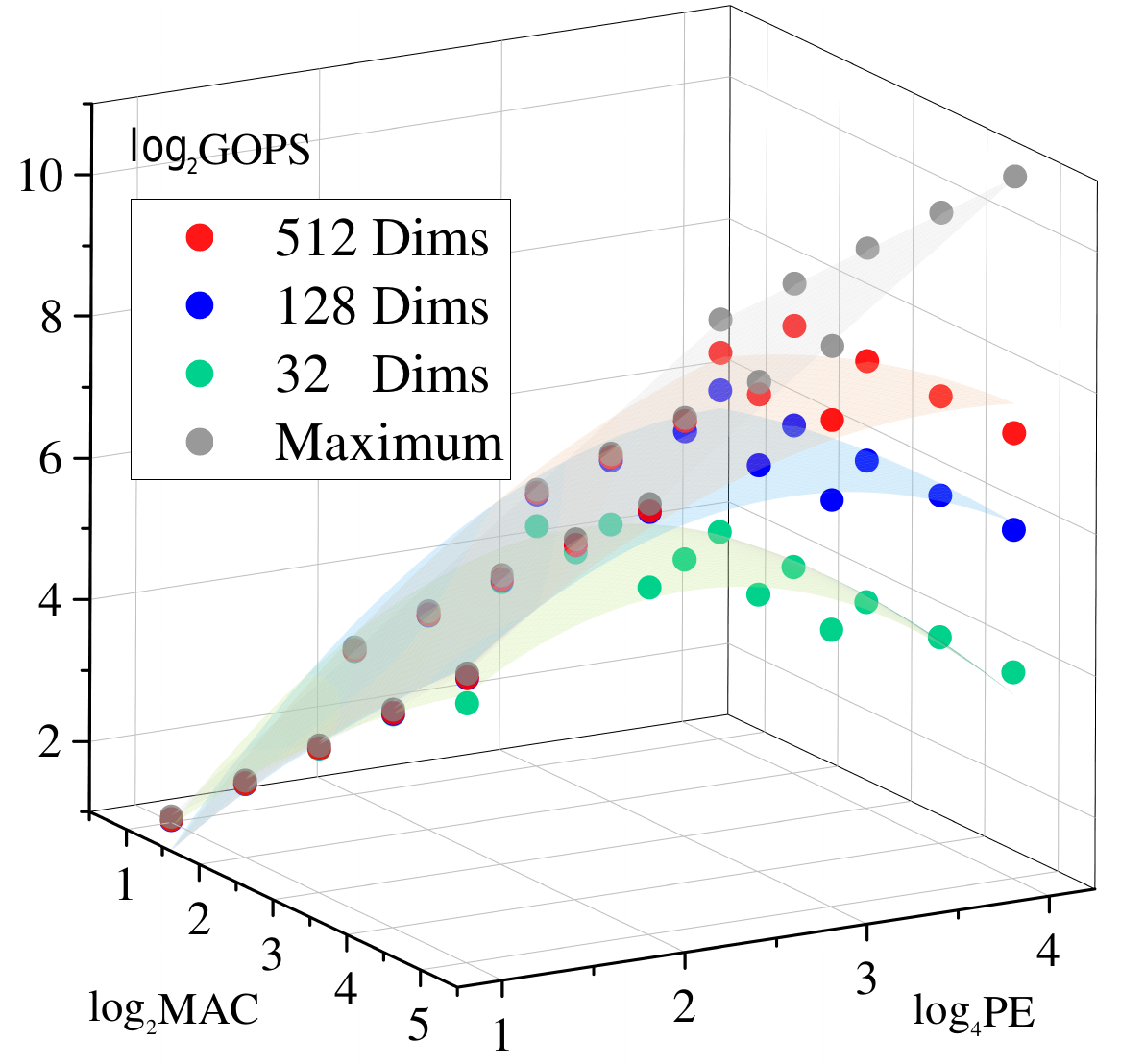}
}
\hspace{-4mm}
\subfigure[Nonlinear Calculation]{
    \includegraphics[trim=0cm 0cm 0cm 0cm, clip,width=0.235\textwidth]{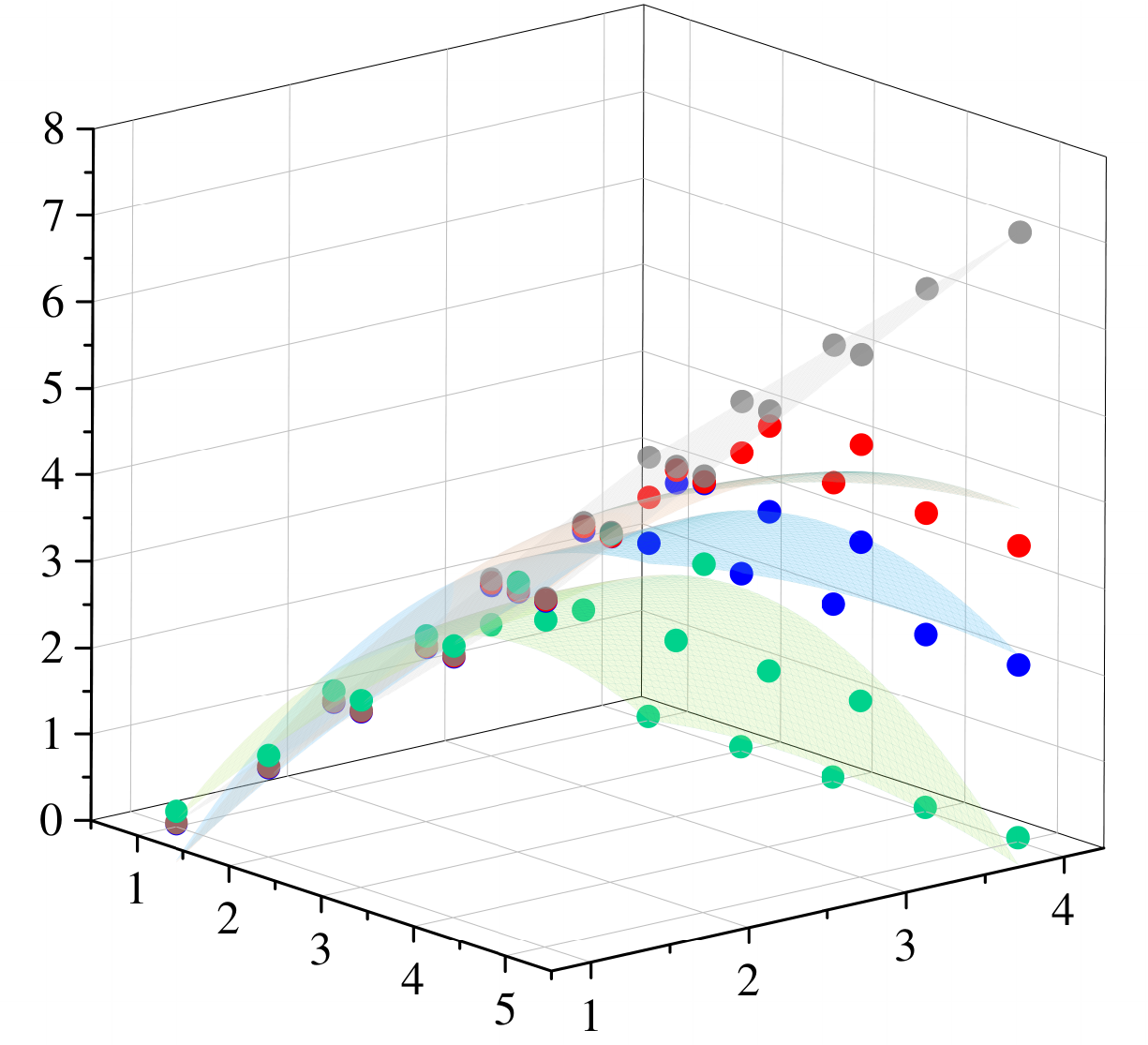}
}

\label{fig:computation_performance}
\caption{Performance under different types of calculation}
\vspace{-8.5mm}

\end{center}
\end{figure}


\subsection{Network Inference Accuracy}
\vspace{-1mm}
Prior to showcasing the computational efficiency and cost of the proposed ONE-SA, we conducted empirical assessments to validate its inference accuracy across popular DNN architecture categories and seventeen diverse tasks. We took diligent measures to ensure that when applying the capped piecewise linearization functions (CPWL) with the appropriate granularity settings, any compromise in accuracy compared to the original models' final performance remained negligible. Table. \ref{tab:accuracy} presents the inference accuracy of various DNN architectures across multiple benchmark datasets. While our experimentation covered 17 benchmark datasets, we limited our presentation to just four benchmarks for each DNN category due to space constraints.
The first column (Original) represents the inference accuracy of the original DNN models with INT16 quantization, which can be considered as the accuracy baselines. The other columns present the inference under capped piecewise linearization with the approximation granularity increasing from 0.1 to 1.0. As expected, the inference accuracy declines as approximation granularity increases since larger granularity incurs higher approximation errors. Regarding specific DNN models, we observe that for both BERT and ResNet, the performance gap between different variants increases as the baseline performance decreases. This suggests that one can choose a larger granularity for easier tasks but a smaller one for more difficult tasks. In contrast, the GCN models do not exhibit significant differences among the baseline and various variants, possibly because GCNs are typically shallower. Theoretically, the proposed ONE-SA architecture can support any approximation granularity. In practice, the approximation granularity is limited by the size of the L3 buffer and the range of uncapped approximation. The granularity of 0.25 is used as our default setting as an illustrative example in the following evaluations. Advanced neural network architecture search (NAS) can also be applied further to select the granularities. 


\begin{figure}[t]
\begin{center}
\setlength{\abovecaptionskip}{-0.02cm}
\subfigure[LUT resources consumption]{
    \includegraphics[trim=0cm 0.4cm 0cm 0cm, clip,width=0.22\textwidth]{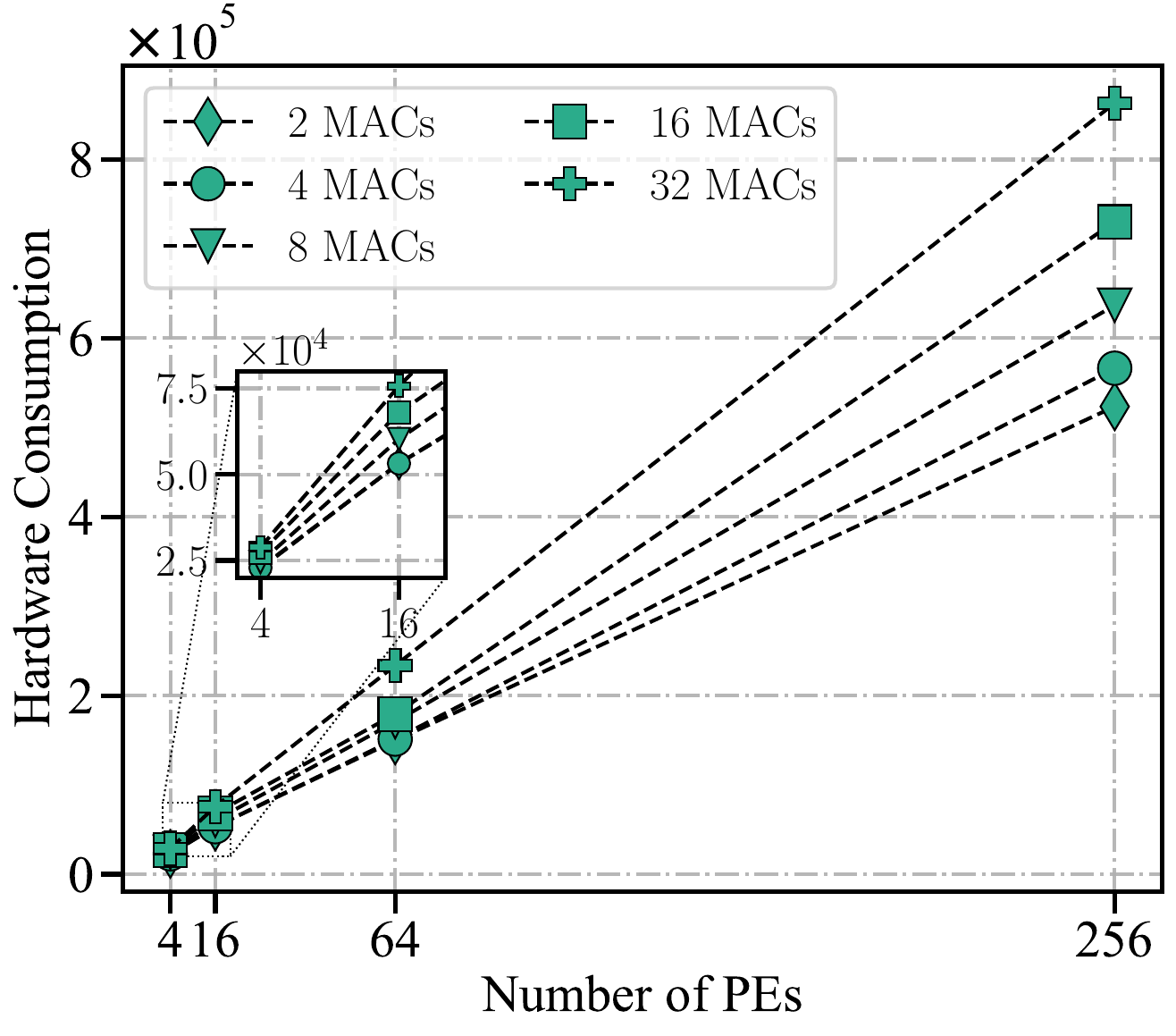}
}
\subfigure[FF resources consumption]{
    \includegraphics[trim=0cm 0.4cm 0cm 0cm, clip,width=0.229\textwidth]{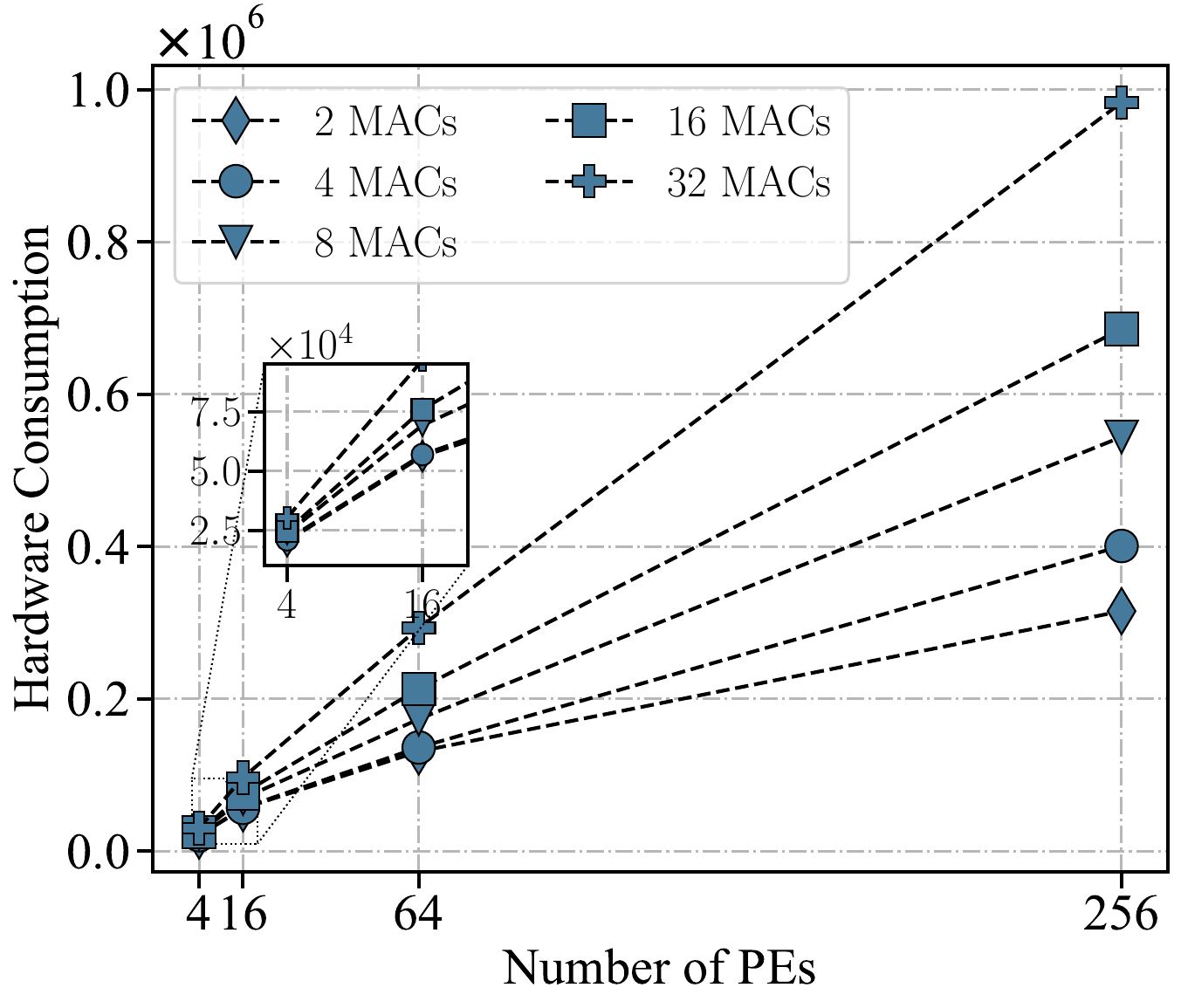}
}
\subfigure[DSP resources consumption]{
    \includegraphics[trim=0cm 0.4cm 0cm 0cm, clip,width=0.22\textwidth]{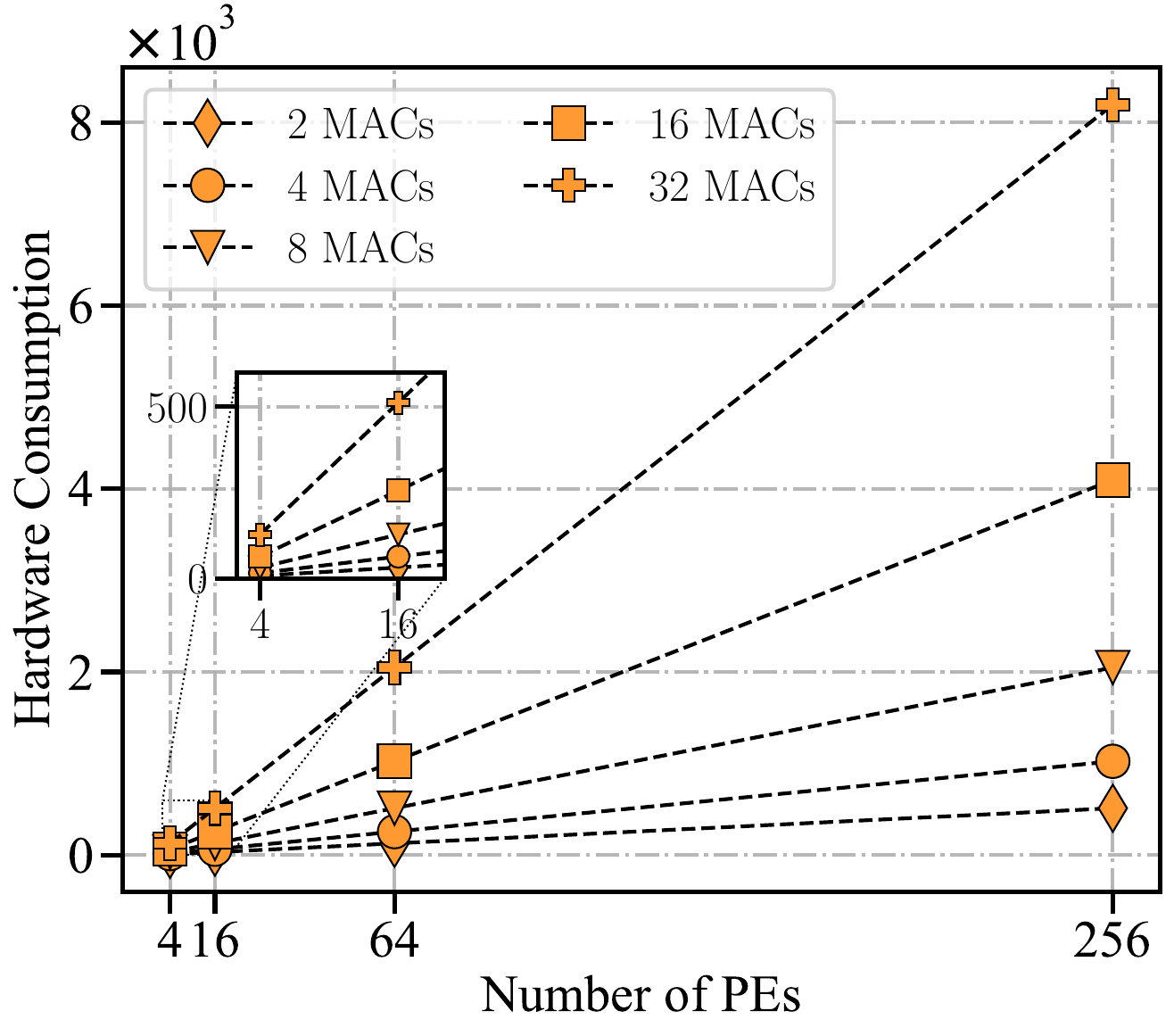}
}
\hspace{-1mm}
\subfigure[BRAM resources consumption]{
    \includegraphics[trim=0cm 0.4cm 0cm 0cm, clip,width=0.229\textwidth]{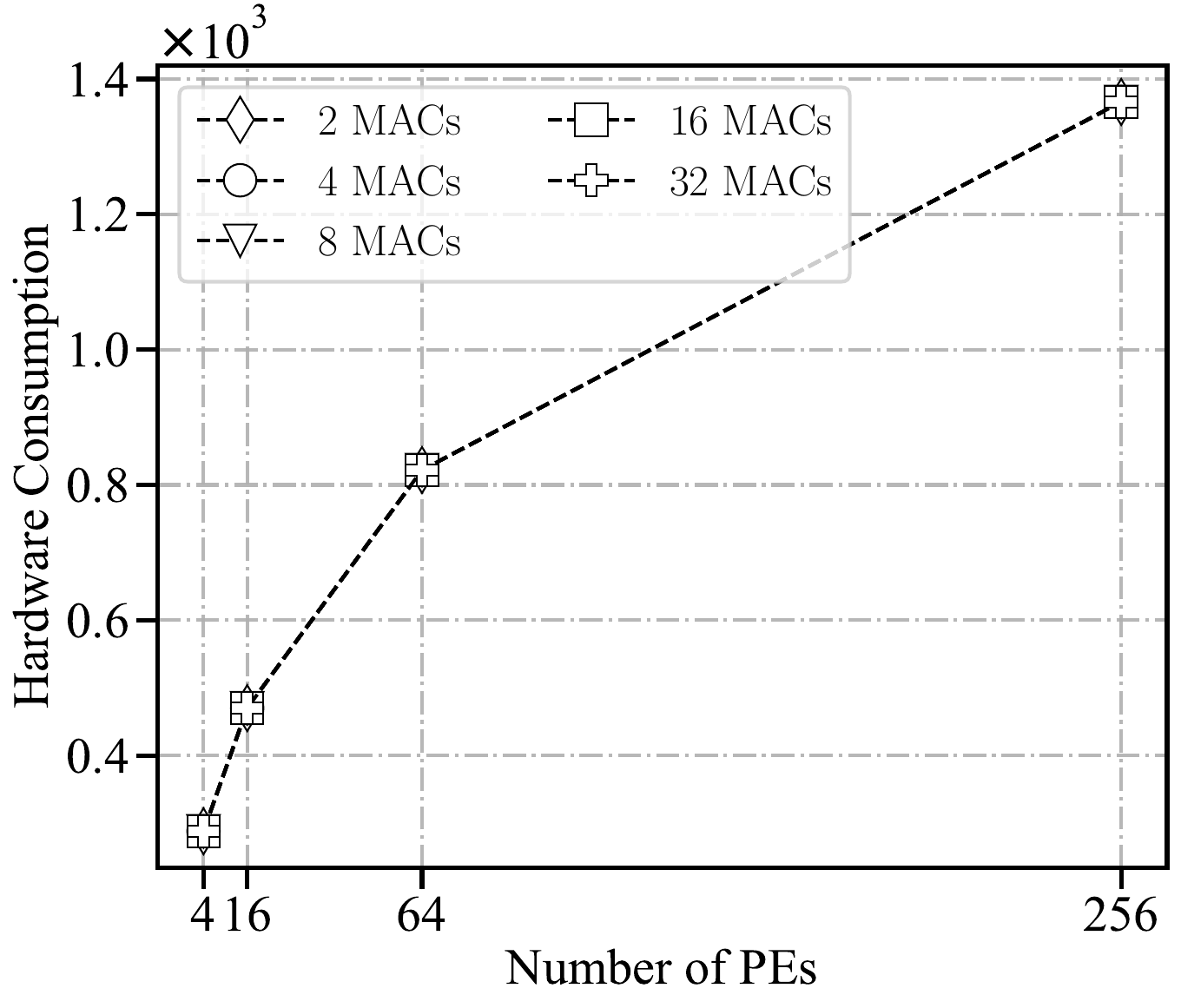}
}

\label{fig:hardware_consumption}
\caption{Resources consumption of ONE-SA with different sizes.}
\vspace{-10mm}
\end{center}
\end{figure}

\vspace{-2mm}
\subsection{Performance and Cost Breakdown}
\vspace{-1mm}
The computational performance of the proposed ONE-SA are presented in Fig. 8. We evaluate the system's performance in both linear computations, quantified in terms of throughput measured in GOPS (Giga Operations Per Second, where each operation encompasses an addition and a multiplication) and nonlinear computations, indicated by the GNFS (Generalized Nonlinear Functionality per Second).
These evaluations encompass a range of scenarios, considering various systolic array sizes, which denote the number of Processing Elements (PEs) and multiply-accumulate units (MACs) within each PE alongside different input matrix sizes. Notably, we observe that both linear and nonlinear computation throughputs exhibit an upward trajectory as the number of PEs and MACs increases, up to a specific point known as the ``throughput cliff."
Moreover, it's worth highlighting that the number of MACs exerts a more pronounced influence on overall throughput compared to other factors.
When compared to the maximum (theoretical) throughput achievable with large matrices, the phenomenon known as the ``throughput cliff" becomes apparent when working with smaller input matrices. For instance, when feeding a small input matrix (e.g. 32x32) into a significantly larger systolic array (e.g. 16x16 PEs), it becomes evident that 84.8\% of the clock cycles are required to transmit the results from the array even after all the computations within the PEs have been completed.
Simultaneously, the proportion of clock cycles dedicated to transmitting inputs to the PEs assumes greater prominence when dealing with small input matrices combined with large systolic arrays. 

\begin{small}
\begin{table*}[!t]
\caption{Performance comparison between ONE-SA and other processors. Different processors are compared by inference Latency (L), relative Speedup (S), Throughput (T), Power (P), and Throughput per Power (T/P).\vspace{-2mm}}
\label{tab:final_comparison}
\resizebox{\linewidth}{!}{
\begin{tabular}{llcrrrrrp{0pt}rrrrrp{0pt}ccccc}

\toprule
\multirow{3}{*}{Processor} & \multirow{3}{*}{Spec} & \multicolumn{1}{l}{\multirow{2}{*}{\begin{tabular}[c]{@{}l@{}}Tech.\\ Node\end{tabular}}} & \multicolumn{5}{c}{ResNet-50} &  & \multicolumn{5}{c}{BERT-base} &  & \multicolumn{5}{c}{GCN} \\ \cline{4-8} \cline{10-14} \cline{16-20} 
 &  & \multicolumn{1}{l}{} & \multicolumn{1}{c}{L} & \multicolumn{1}{c}{S} & \multicolumn{1}{c}{T} & \multicolumn{1}{c}{P} & \multicolumn{1}{c}{T/P} & \multicolumn{1}{c}{} & \multicolumn{1}{c}{L} & \multicolumn{1}{c}{S} & \multicolumn{1}{c}{T} & \multicolumn{1}{c}{P} & \multicolumn{1}{c}{T/P} & \multicolumn{1}{c}{} & L & S & T & P & T/P \\
 &  & \multicolumn{1}{c}{(nm)} & \multicolumn{1}{c}{(ms)} & \multicolumn{1}{c}{(x)} & \multicolumn{1}{c}{(GOPS)} & \multicolumn{1}{c}{(W)} & \multicolumn{1}{c}{(1/W)} & \multicolumn{1}{c}{} & \multicolumn{1}{c}{(ms)} & \multicolumn{1}{c}{(x)} & \multicolumn{1}{c}{(GOPS)} & \multicolumn{1}{c}{(W)} & \multicolumn{1}{c}{(1/W)} & \multicolumn{1}{c}{} & (ms) & (x) & (GOPS) & (W) & (1/W) \\ \midrule
Intel CPU & i7-11700 & 14 & 42.51 & 1.00 & 93.51 & 112.0 & \textbf{0.83} &  & 45.92 & 1.00 & 119.77 & 112.0 & \textbf{1.07} &  & \multicolumn{1}{r}{34.12} & \multicolumn{1}{r}{1.00} & \multicolumn{1}{r}{33.99} & \multicolumn{1}{r}{112.0} & \multicolumn{1}{r}{\textbf{0.30}} \\
NVIDIA GPU & 3090Ti & 8 & 6.27 & 6.78 & 633.99 & 131.0 & \textbf{4.84} &  & 7.95 & 5.78 & 691.81 & 131.0 & \textbf{5.28} &  & \multicolumn{1}{r}{1.56} & \multicolumn{1}{r}{21.87} & \multicolumn{1}{r}{743.45} & \multicolumn{1}{r}{131.0} & \multicolumn{1}{r}{\textbf{5.68}} \\
NVIDIA SoC & AGX ORIN & 12 & 16.20 & 2.62 & 245.38 & 14.0 & \textbf{17.53} &  & 21.52 & 2.13 & 255.57 & 14.0 & \textbf{18.26} & \multicolumn{1}{r}{} & \multicolumn{1}{r}{4.92} & \multicolumn{1}{r}{6.930} & \multicolumn{1}{r}{235.73} & \multicolumn{1}{r}{14.0} & \multicolumn{1}{r}{\textbf{16.84}} \\ \midrule
Zynq Z-7020 & Angel-eye\cite{7930521} & 28 & 47.15 & 0.90 & 84.3 & 3.5 & \textbf{24.09} &  & \multicolumn{1}{c}{-} & \multicolumn{1}{c}{-} & \multicolumn{1}{c}{-} & \multicolumn{1}{c}{-} & \multicolumn{1}{c}{-} &  & - & - & - & - & - \\
Virtex7 & VGG16 \cite{8309067} & 28& 19.64 & 2.16 & 202.42 & 10.81 & \textbf{18.72} &  & \multicolumn{1}{c}{-} & \multicolumn{1}{c}{-} & \multicolumn{1}{c}{-} & \multicolumn{1}{c}{-} & \multicolumn{1}{c}{-} &  & - & - & - & - & - \\
Zynq Z-7100 & NPE\cite{khan2021npe} & 28 & \multicolumn{1}{c}{-} & \multicolumn{1}{c}{-} & \multicolumn{1}{c}{-} & \multicolumn{1}{c}{-} & \multicolumn{1}{c}{-} &  & 13.57 & 3.38 & 405.30 & 20.0 & \textbf{20.27} &  & - & - & - & - & - \\
Virtex UltraScale+ & FTRANS\cite{li2020ftrans} & 16 & \multicolumn{1}{c}{-} & \multicolumn{1}{c}{-} & \multicolumn{1}{c}{-} & \multicolumn{1}{c}{-} & \multicolumn{1}{c}{-} &  & 9.82 & 4.68 & 559.85 & 25 & \textbf{22.39} &  & - & - & - & - & - \\ \midrule
\textbf{Virtex7} & ONE-SA & 28 & \textbf{26} & \textbf{1.64} & \textbf{152.89} & \textbf{7.61} & \textbf{20.09} &  & \textbf{26.24} & \textbf{1.75} & \textbf{209.60} & \textbf{7.61} & \textbf{27.54} &  & \multicolumn{1}{r}{\textbf{5.87}} & \multicolumn{1}{r}{\textbf{5.81}} & \multicolumn{1}{r}{\textbf{197.58}} & \multicolumn{1}{r}{\textbf{7.61}} & \multicolumn{1}{r}{\textbf{25.96}} \\ \bottomrule

\end{tabular}
}
\vspace*{-5mm}
\end{table*}
\end{small}

The resources cost associated with various systolic array sizes are succinctly presented in Fig. 9. As the number of PEs increases, there is a discernible linear growth in the utilization of LUTs, FFs, and DSPs. In contrast, the consumption of BRAM experiences a more gradual increment. This observation underscores that a significant proportion of LUTs, FFs, and DSPs resources are primarily allocated to support the PEs within the systolic array. Furthermore, when the number of MACs within each PE is increased, there is a linear rise in DSPs resource utilization. The utilization of FFs increases by approximately 2.6\% to 53.8\% when double the number of MACs is employed. However, the most noteworthy finding pertains to the consumption of LUTs and BRAMs. The utilization of LUTs sees a marginal increase and there is no corresponding increase in BRAM utilization with the additional MACs. In conjunction with the aforementioned throughput performance, it becomes evident that an increase in the number of MACs leads to higher throughput while incurring a relatively smaller resource overhead.

The execution time and power consumption of different input matrix dimensions under different systolic array design sizes are presented in Fig. 10, where all the design points are scatter-plotted, and the Pareto frontiers consist of the optimal design points. We distinguish the designs by the number of MACs in each PE and mark them with different colors. Clearly, the designs with more MACs will have a better performance (lower latency and lower power consumption). The designs with 16 or more MACs are closely located at the Pareto frontiers, which indicates that 16-MAC are an optimal design choice, and adding more MACs will not effectively push the Pareto frontiers to a lower compulsion and latency. Meanwhile, the optimal design points for classic linear computation (i.e., general matrix multiply) are also the optimal or near-optimal designs for the new-enabled nonlinear computation. \vspace{-4mm}

\begin{figure}[tbp]
\begin{center}

\setlength{\abovecaptionskip}{-0.1cm}
\subfigure[Linear computation]{
    \includegraphics[trim=0cm 0cm 0cm 0cm, clip,width=0.23\textwidth]{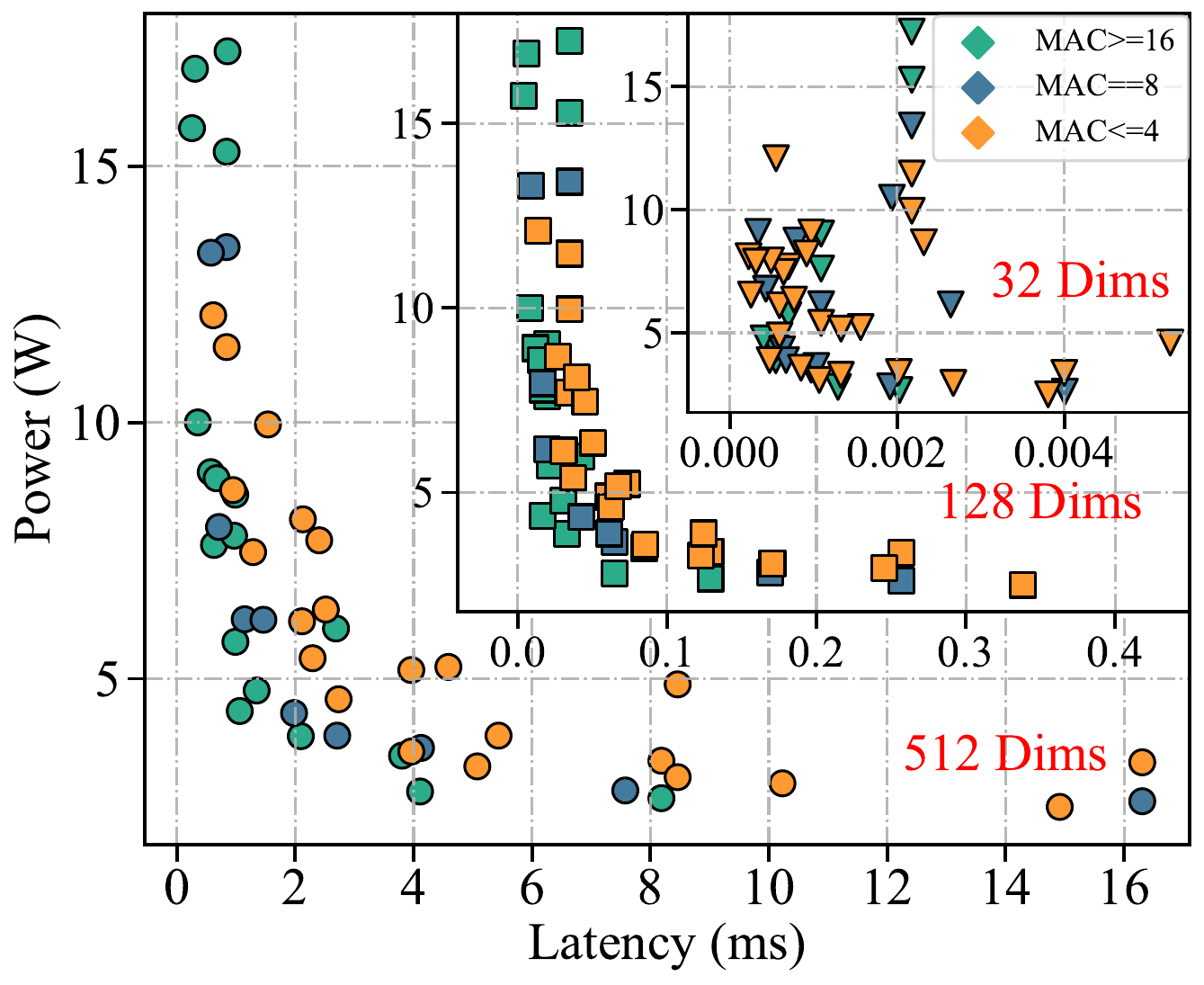}
}
\hspace{-4mm}
\subfigure[Nonlinear computation]{
    \includegraphics[trim=0cm 0cm 0cm 0cm, clip,width=0.234\textwidth]{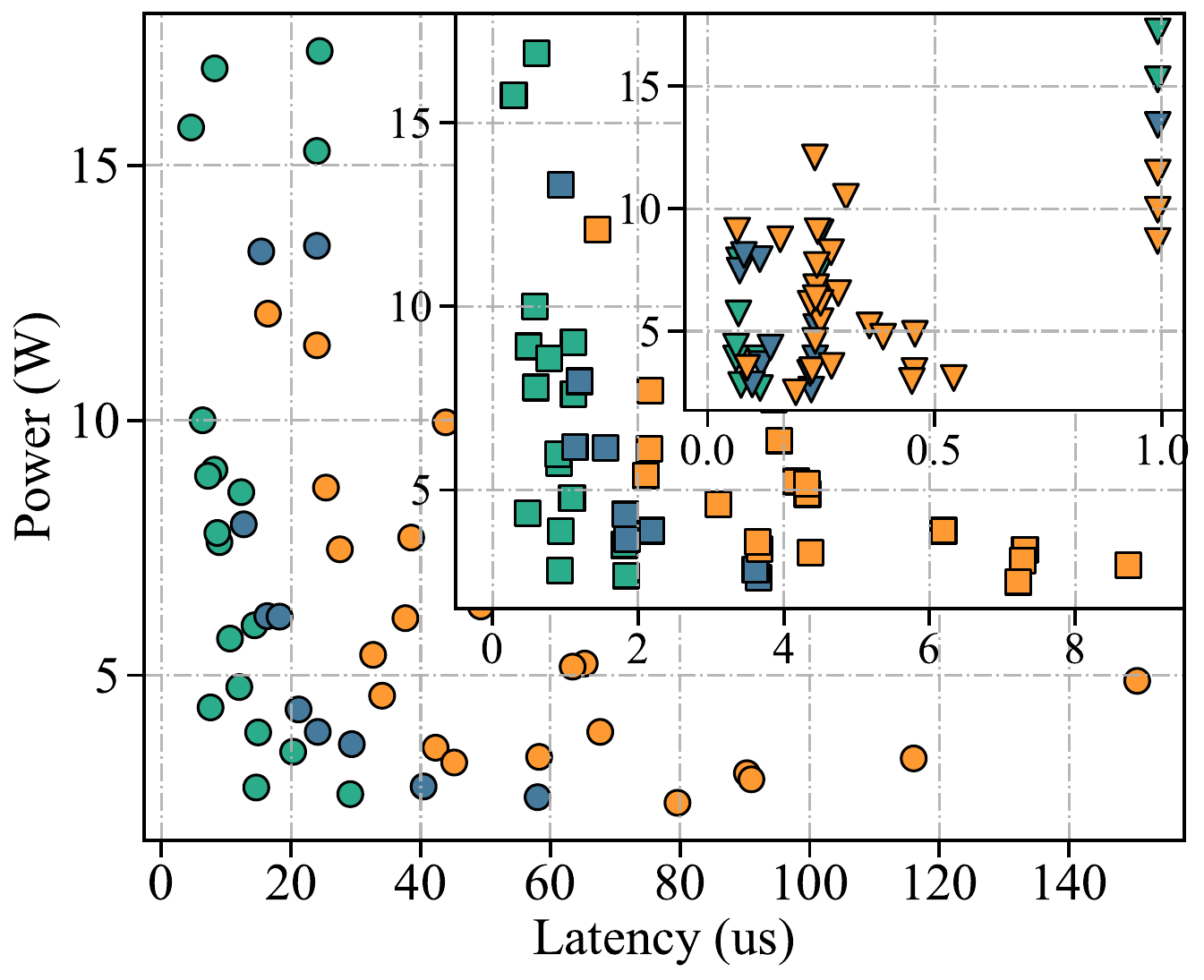}
}

\label{fig:latency_power}
\caption{Computation latency with power consumption.}
\vspace{-8mm}

\end{center}
\end{figure}

\begin{table}[h]\scriptsize
  \centering
  \caption{Buffer sizes\vspace{-1mm}}
    \begin{tabular}{c|cccc}
    \hline
    Buffers in & L3    & L2    & PE    & L1 \\
    \hline
    Size  & 0.28KB & 0.5KB & 0.094KB & 0.031KB \\
    Total & 0.28KB$\times$3 & 0.5KB$\times$24 & 0.094KB$\times$64 & 0.031KB$\times$64 \\
    \hline
    \end{tabular}%
  \label{tab:addlabel}%
  \vspace{-6mm}
\end{table}%

\subsection{Comparison with General-Purpose and App.-Specific Processors}
\vspace{-1mm}
Finally, the ONE-SA (64 PEs, 16 MACs in each PE, the sizes of each buffer is presented in Table ) is compared with general-purpose processors CPU, GPU, SoC and the application-specific accelerators implemented by the FPGAs. The latency and throughput of general-purpose processors can be directly logged from the operating system. The power consumption is measured with a TCP2020 current probe and Tektronix MDO32 oscilloscope. To compare the proposed ONE-SA with other application-specific accelerator designs, we pick the recent FPGA-based accelerator designs for the ResNet-50 \cite{7930521, 8309067} and BERT \cite{khan2021npe} (or RoBERTa \cite{li2020ftrans} same architecture with BERT). For the GNN, we can only present performance data for runs on general-purpose processors, as we were unable to locate standard ASIC designs for GNN, which has numerous variations. 

To standardize the performance with the power consumption in different accelerators, the throughput per power is regarded as the metric for computation efficiency. 
Moving ResNet-50, BERT, and GCN to the proposed ONE-SA achieves up to 86.53$\times$ and 5.21$\times$ improvements on computation efficiency compared with the general-purpose CPU and GPU. 
For the embedded SoCs, up to 1.54$\times$ improvements in computation efficiency can be achieved in running the above neural network.
Compared with the ResNet-specific and BERT-specific FPGA designs, the same levels (i.e. 83.40\% - 135.87\%) of computation efficiency are achieved together with the flexibility of running versatile neural networks on one systolic array accelerator. \vspace{-2mm}

%% file: Conclusion.tex
\section{Conclusion}
\label{sec:Conclusion}
\vspace{-2mm}
This paper presents ONE-SA, a lightweight systolic array architecture support for nonlinear operations. Utilizing capped piecewise functions (CPWL), the nonlinear operations are achieved through Intermediate Parameter Fetching and Matrix Hadamard Product. To facilitate these two architecture events in the systolic array, efficient data paths are designed and optimized within the L3 buffer and the PEs. These data paths support the efficient delivery of intermediate parameters to the PEs and the following data movements within PEs. Experiments demonstrate that ONE-SA achieves efficiency levels comparable to those of ASIC-based accelerators while retaining the flexibility to accommodate versatile neural network models.
\vspace{-2mm}